\documentclass[aip,superscriptaddress,reprint]{revtex4-2}

\input{preamble.sty}

\begin{document}

\title{Robust quantification of the diamond nitrogen-vacancy center charge state via photoluminescence spectroscopy}

\author{G. Thalassinos}
\email{giannis.thalassinos@rmit.edu.au}
\affiliation{School of Science, RMIT University, Melbourne, VIC 3001, Australia}

\author{D. J. McCloskey}
\affiliation{School of Physics, University of Melbourne, Parkville, VIC 3010, Australia}
\affiliation{Australian Research Council Centre of Excellence in Quantum Biotechnology, School of Physics, University of Melbourne, Parkville, VIC 3010, Australia}

\author{A. Mameli}
\affiliation{School of Science, RMIT University, Melbourne, VIC 3001, Australia}

\author{A. J. Healey}
\affiliation{School of Science, RMIT University, Melbourne, VIC 3001, Australia}

\author{C. Pattinson}
\affiliation{School of Physics, University of Melbourne, Parkville, VIC 3010, Australia}
\affiliation{Australian Research Council Centre of Excellence in Quantum Biotechnology, School of Physics, University of Melbourne, Parkville, VIC 3010, Australia}

\author{D. Simpson}
\affiliation{School of Physics, University of Melbourne, Parkville, VIC 3010, Australia}
\affiliation{Australian Research Council Centre of Excellence in Quantum Biotechnology, School of Physics, University of Melbourne, Parkville, VIC 3010, Australia}

\author{B. C. Gibson}
\affiliation{School of Science, RMIT University, Melbourne, VIC 3001, Australia}

\author{A. Stacey}
\affiliation{School of Science, RMIT University, Melbourne, VIC 3001, Australia}

\author{N. Dontschuk}
\email{dontschuk.n@unimelb.edu.au}
\affiliation{School of Physics, University of Melbourne, Parkville, VIC 3010, Australia}

\author{P. Reineck}
\email{philipp.reineck@rmit.edu.au}
\affiliation{School of Science, RMIT University, Melbourne, VIC 3001, Australia}

 
\begin{abstract}
\noindent 
Nitrogen vacancy (NV) centers in diamond are at the heart of many emerging quantum technologies, all of which require control over the NV charge state. 
Hence, methods for quantification of the relative photoluminescence (PL) intensities of the  \nvn{} and \nvm{} charge state, i.e., a charge state ratio, are vital. 
Several approaches to quantify NV charge state ratios have been reported but are either limited to bulk-like NV diamond samples or yield qualitative results.
We propose an NV charge state quantification protocol based on the determination of sample- and experimental setup-specific \nvn{} and \nvm{} reference spectra. 
The approach employs blue (\numrange{400}{470}~nm) and green (\numrange{480}{570}~nm) excitation to infer pure \nvn{} and \nvm{} spectra, which are then used to quantify NV charge state ratios in subsequent experiments via least squares fitting. 
We test our dual excitation protocol (DEP) for a bulk diamond NV sample, 20 and 100~nm nanodiamond particles and compare results with those obtained via other commonly used techniques such as zero-phonon line fitting and non-negative matrix factorization.  
We find that DEP can be employed across different samples and experimental setups and yields consistent and quantitative results for NV charge state ratios that are in agreement with our understanding of NV photophysics. 
By providing robust NV charge state quantification across sample types and measurement platforms, DEP will support the development of NV-based quantum technologies.

\end{abstract}
\noindent

\maketitle
\section{Introduction}
\noindent
The nitrogen vacancy (NV) center in diamond is one of the most widely adopted platforms for emerging quantum technologies~\cite{schirhagl2014nitrogenvacancy,pezzagna2021quantum,wrachtrup2006processing}.
In bulk diamond, NV centers are generally present in either a neutral (\nvn) or a negative charge state (\nvm). 
Although both charge states are photoluminescent, only \nvm{} has an optically addressable spin that allows for spin-based optical quantum sensing, enabling applications including magnetometry~\cite{rondin2014magnetometry} and thermometry~\cite{neumann2013highprecision}.
Furthermore, the NV charge state itself can be used as an all-optical nanoscale voltage~\cite{mccloskey2022diamond} and electric-field sensor~\cite{dolde2011electricfield}.
Hence, significant materials engineering~\cite{doi2016pure} and photonics~\cite{wood2024wavelength,barry2020sensitivity} efforts are underway to control and tailor the NV charge state for specific applications.

\begin{figure}
    \centering
    \includegraphics[width=0.48\textwidth]{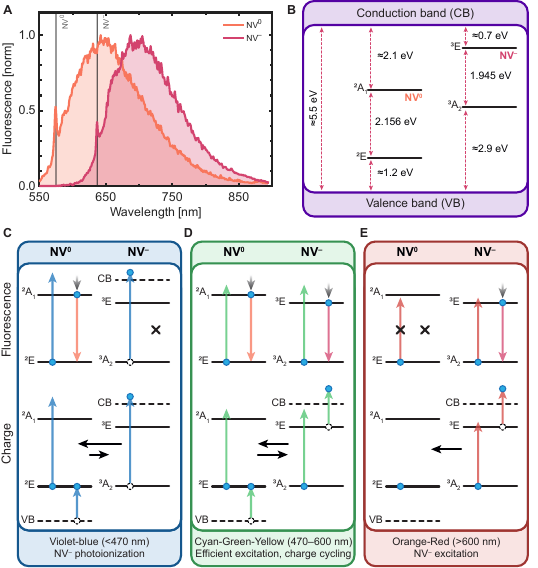}
    \caption{NV center photodynamics.
    \textbf{A:}~Fluorescence spectra of \nvn{} and \nvm{} with characteristic ZPLs at 575 and 637~nm (marked via vertical lines), respectively, showing significant overlap between \numrange{600}{850}~nm.
    \textbf{B:}~Energy level diagram of \nvn{} and \nvm{} within the diamond band gap.
    \textbf{C-E:}~Dynamics of the NV center under different excitation wavelengths.
    Violet-blue excitation (C) leads to single-photon photoionization of \nvm{} into \nvn, while still exciting \nvn, which relaxes down to the ground state via fluorescence.
    Green illumination (D) leads to efficient excitation of both charge states and allows for dynamic charge state switching via ionization and re-pumping processes.
    Red light (E) has insufficient energy to promote \nvn{} to the excited state, leading exclusively to \nvm{} emission.
    Horizontal, black lines indicate charge state switching, with relative length indicative of the rate in a particular direction.
    Vertical, short grey lines represent phonon relaxations.
    }
    \label{fig:energy_levels}
\end{figure}

Underpinning all of these efforts is the ability to quantify the NV charge state. 
While the two charge states can be clearly identified in PL spectra via their zero-phonon lines (ZPLs) at 575.07~nm (2.156~eV, \nvn) and 637.45~nm (1.945~eV, \nvm), their phonon side-bands (PSBs) overlap significantly (\cref{fig:energy_levels}A)~\cite{rondin2010surfaceinduced,manson2018nv,aslam2013photoinduced}, making precise quantification of the relative contribution of each charge state to the overall NV PL non-trivial. 
Several approaches for the quantification of the NV charge state have been reported. One approach uses the intensity ratio of the two ZPLs as a measure of the relative contribution of each charge state to the overall PL signal~\cite{acosta2009diamonds,alsid2019photoluminescence,fu2010conversion,shinei2021equilibrium}. 
This approach is feasible for NV ensemble measurements in bulk diamond samples, where pronounced NV ZPLs are generally observed, particularly at low temperatures~\cite{manson2005photoionization}. 
However, for NV centers in small nanodiamonds~\cite{eldemrdash2023fluorescent,rondin2010surfaceinduced,bradac2010observation} and, more generally, for NV centers within a few nanometers of the diamond surface~\cite{mccloskey2022diamond}, ZPLs are weak or not visible at all in PL spectra. 
Thus, another approach to quantify the NV charge state ratio is based on the use of `pure' \nvn{} and \nvm{} spectra from literature such as reference~\cite{rondin2010surfaceinduced} that are used as fits in other laboratories~\cite{shenderova2017commercial,groot-berning2014passive,dhomkar2018charge}.
However, NV spectral profiles can vary depending on the diamond properties (e.g., strain), sample dimensions (bulk vs. nanoparticles), environmental factors like temperature~\cite{doherty2014temperature,audecraik2020microwaveassisted}, the nanoscale environment~\cite{zhao2012suppression}, and the spectroscopy equipment used. 
As a result, the seemingly simple question `How far into the far-red do \nvn{} and \nvm{} emit?' has no simple answer based on the current literature (see SI Table S1).
Hence, pure \nvn{} and \nvm{} spectra are ideally acquired in a sample- and equipment-specific manner.

To this end, three main approaches have been reported. 
The first two approaches are photoluminescence (PL) decomposition analysis~\cite{alsid2019photoluminescence} and a similar microwave-assisted decomposition technique~\cite{audecraik2020microwaveassisted}. 
The former modulates the NV charge state via the excitation power to change the \nvm{} ionization rate; the latter modulates the \nvm{} PL intensity using microwaves. 
Both techniques use their respective modulation to first derive the spectral profile of \nvm.
Then, this \nvm{} reference is subtracted from a mixed \nvn/\nvm{} signal until there is no visible ZPL peak at 637~nm to determine \nvn.
The third approach employs machine learning which does not require \textit{a-priori} knowledge about the NV ZPLs in order to derive a set of pure NV spectra. 
Non-negative matrix factorization (NNMF) is an often used algorithm for decomposition of spectra into an integer number of non-negative components for a given set of training data. The training set consists of a series of NV spectra, where each spectrum contains different relative contributions of \nvn{} and \nvm~\cite{reineck2019not,wilson2019effect,mccloskey2020enhanced,savinov2022diamond}. 
However, it is not clear that NNMF accurately determines the ``true'' \nvn/\nvm{} spectral profiles.
Compared to experiment~\cite{manson2018nv,manson2005photoionization,alsid2019photoluminescence,groot-berning2014passive,aslam2013photoinduced,audecraik2020microwaveassisted,bhaumik2019tunable,rondin2010surfaceinduced,treussart2006photoluminescence,karaveli2016modulation}, NNMF predicts that the maximum emission wavelength of \nvn{} is significantly shorter than \nvm{} by $>$70~nm (see SI Table S1)~\cite{mccloskey2020enhanced,reineck2019not,wilson2019effect,savinov2022diamond} and can show the presence of the \nvm{} ZPL in the `pure' \nvn{} spectrum and/or vice-versa~\cite{mccloskey2020enhanced}. 
Therefore, new approaches are needed.

Here, we propose a dual wavelength excitation protocol (DEP) for measuring the pure \nvn{} and \nvm{} spectral profiles, which are used to determine the NV charge state ratio. 
We show that this method works across a wide variety of diamond types, including bulk diamond and nanoparticles; and vastly different optical configurations, including confocal microscopes and nanoparticles in suspension. 
We leverage the direct photo-ionization of \nvm{} using blue/violet excitation ($<$470~nm) to obtain a pure \nvn{} spectrum, which is then used to determine the \nvm{} spectrum. 
\nvn{} and \nvm{} reference spectra are then fit to experimental data using the linear least-squares fitting algorithm to determine the NV charge state ratio. 
We benchmark this technique against ZPL fitting, NNMF, and skewed Gaussian (SG) fitting to approximate the `pure' NV spectra. 
We demonstrate that our protocol works across a range of samples and measurement techniques, even in the absence of ZPLs. 
Our protocol enables robust quantification of the NV charge state and will thereby aid the development of next-generation quantum sensing materials and protocols.  

\section{Methods}

\subsection{Confocal fluorescence spectroscopy}
We investigated a commercially available bulk diamond sample (DNV-B14, Element Six) and fluorescent nanodiamonds (FNDs) with a nominal diameter of 100~nm (Adàmas Nanotechnologies) spin-coated onto a Si substrate using our custom-built scanning confocal microscope (SI fig. S1A). 
Samples were excited using an 80~MHz pulsed laser (Fianium WhiteLase 400-SC, NKT Photonics) operating between \numrange{400}{700}~nm through a 100$\times$ objective (NA~=~0.9).
Fluorescence was collected through the same objective and fibre coupled to an avalanche photodiode (APD) (25~\%, SPCM-AQRH-14-TR, Excelitas) and a spectrometer (75~\%, SpectraPro SP-2500, Princeton Instruments) with a charge-coupled device (CCD) camera (PIXIS 100BR, Princeton Instruments).

\subsection{In-solution fluorescence spectroscopy} 
Fluorescent nanodiamonds (FNDs) with nominal sizes of 20 and 140~nm (Adámas Nanotechnologies) were suspended in de-ionized (DI) water at approximately 1.00 and 0.01~\si{\mg\per\mL}, respectively before being measured in our custom-built in-solution spectroscopy setup (SI fig. S1B).  
Samples were excited using the same pulsed laser as the confocal setup and weakly focused onto the sample cuvette (18FL, FireFlySci). 
Fluorescence was collected via a custom fibre (\si{105~\um} core, NA~=~0.22) through an in-line filter mount (FOFMS, Thorlabs) and coupled into the spectrometer. 
For each excitation wavelength, a water Raman spectrum and a scattering reference using 40~nm SiO$_2$ nanoparticles were acquired and subtracted from the FND signals (see SI for details).

\section{Dual excitation protocol}
Generally, accurately determining the spectral profile of \nvm{} can be achieved using any technique which requires at least two measured spectra with a different NV charge state ratio~\cite{alsid2019photoluminescence,wilson2019effect,savinov2022diamond,audecraik2020microwaveassisted,reineck2019not}.
The fundamental challenge lies in the robust determination of the \nvn{} spectral profile in the spectral region above 700~nm, where it significantly overlaps with \nvm. 

The \nvm{} ground state is located $\approx$2.6~eV below the conduction band~\cite{aslam2013photoinduced} (\cref{fig:energy_levels}B). 
It is efficiently photoionized to \nvn{} via a single-photon processes using excitation wavelengths below $\approx$470~nm, leading to \nvn{} emission.  
In practice, it has been observed that NV PL spectra acquired using excitation wavelengths of 458~nm~\cite{manson2018nv,yang2022photoluminescence} still show some \nvm{} PL in bulk diamond, which disappears for excitation wavelengths $<$450~nm. 
Under green excitation (\numrange{510}{540}~nm), both NV charge states are efficiently excited and cycle between each other through ionization and recombination processes (\cref{fig:energy_levels}D)~\cite{aslam2013photoinduced}.
Under red excitation (\numrange{600}{640}~nm), photons only have sufficient energy to excite \nvm~(\cref{fig:energy_levels}E).
Simply, there exist excitation windows in which either one of the two NV charge states is directly excited, or both are.
Therefore, by tuning the excitation wavelength, sample-specific reference \nvn/\nvm{} spectra can be obtained.

\begin{figure}
\centering
\includegraphics[width=0.48\textwidth]{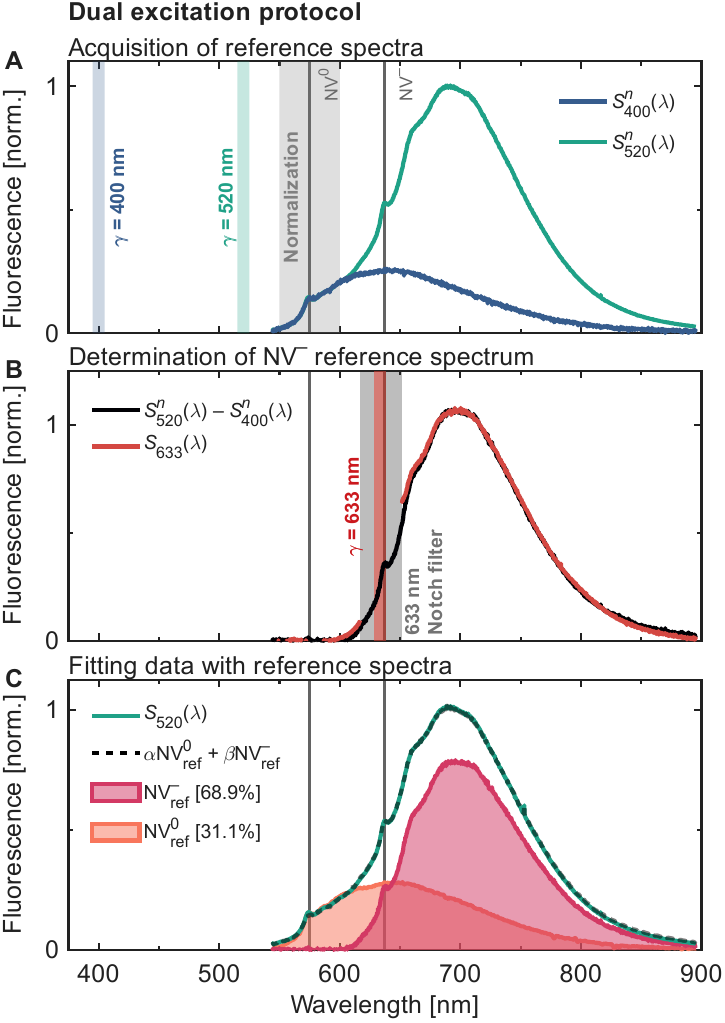}
\caption{Dual excitation protocol on 140~nm FNDs suspended in water.
\textbf{A:}~Reference spectra are acquired using 400 and 520~nm excitation to obtain a pure \nvn{} signal ($S_{400}^n(\lambda)$) and a mixed \nvn/\nvm{} spectrum ($S_{520}^n(\lambda)$), respectively, which are then normalized for the same \nvn{} contribution between \numrange{550}{600}~nm (shaded region).
\textbf{B:}~A pure \nvm{} spectrum is obtained by taking the difference between the two reference spectra in (A), which is then compared to experiment under 633~nm illumination ($S_{633}(\lambda)$). 
\textbf{C:}~The reference \nvn{} and \nvm{} spectra can then be used to fit experimental data using \cref{eq:nv_fitting} to determine the charge state ratio. 
}
\label{fig:workflow_results}
\end{figure}

\Cref{fig:workflow_results} illustrates the main steps of our protocol based on measurements performed on commercially available fluorescent nanodiamonds (FNDs) with a nominal size of 140~nm suspended in DI water.  
Pure \nvn{} and mixed \nvn/\nvm{} fluorescence spectra were acquired using violet (400~nm) and green (520~nm) excitation, respectively, as shown in \cref{fig:workflow_results}A. 
The measured spectra are labelled $S_\gamma(\lambda)$ where $\gamma$ and $\lambda$ represent the excitation and emission wavelengths, respectively.

Each measured NV signal can be expressed as a linear combination of the two charge states 

\begin{equation}\label{eq:nv_fitting}
    S_{\gamma}(\lambda) = \alpha(\gamma)\text{NV}^0 + \beta(\gamma)\text{NV}^{‒},
\end{equation}

where $\alpha(\gamma)$ and $\beta(\gamma)$ are unknown non-negative scalar quantities representing the relative contribution of each charge state as a function of the excitation wavelength.
In the case where $\gamma$~$<$470~nm, $\alpha$~$>$~0 and we assume $\beta~\approx$~0, yielding us a reference \nvn{} spectrum (\nvn$_\text{ref}$),

\begin{equation}
    S_{<470}(\lambda) = \alpha(<470)\text{NV}^{0} = \text{NV}^0_\text{ref}.
\end{equation}

\nvm{} exhibits negligible fluorescence below 600~nm~\cite{manson2018nv}, thus both spectra were normalized (denoted by superscript $n$) between $\lambda$~=~\numrange{550}{600}~nm (highlighted in grey in \cref{fig:workflow_results}A), via the expression 

\begin{equation}
S_{\gamma}^{n}(\lambda) = \frac{S_{\gamma}(\lambda)}{\int_{550}^{600} S_{\gamma}(\lambda) \,d\lambda},
\end{equation}

thus, both signals are normalized to the same \nvn{} intensity, or

\begin{equation}
    \frac{\alpha(<470)}{\alpha(520)} = 1.
\end{equation}

This normalization allows us to fully subtract the \nvn{} signal from any mixed \nvn/\nvm{} spectrum, resulting in a pure \nvm{} reference spectrum (\nvm$_\text{ref}$), via the relation

\begin{equation}
    \text{NV}^-_\text{ref} = S_{470-600}^{n}(\lambda) - S_{<470}^{n}(\lambda),
\end{equation}

as seen in \cref{fig:workflow_results}B.
Here, we see that the \nvm{} ZPL correctly appears at 637~nm, with no sign of \nvn{} fluorescence.
For verification, we investigated the same sample under 633~nm (1.959~eV) excitation, which only has sufficient energy to excite \nvm, labeled ($S_{633}(\lambda)$). We found that red excitation leads to no detectable \nvn{} signal and is nearly identical to the DEP derived \nvm{} spectrum.

With these two measurements, we can fit any combination of \nvn{} and \nvm{} using the least squares method to determine the relative contribution of each charge state and precisely determine the \nvm{} contribution via the expression

\begin{equation}
\small
    \text{NV}^- = \frac{\beta(\gamma)  \int\limits_{\lambda=0}^{\inf} \text{NV}^-_{\text{ref}}(\lambda) \,d\lambda}
    {\int\limits_{\lambda=0}^{\inf} \alpha(\gamma)  \text{NV}^0_{\text{ref}}(\lambda) + \beta(\gamma) \text{NV}^-_{\text{ref}}(\lambda) \,d\lambda  }.
\end{equation}

An example is provided in \cref{fig:workflow_results}C, where the two reference spectra are used to fit the original green data ($S_{520}(\lambda)$), revealing an \nvm{} contribution of 68.9~\%. 

\subsection{Benchmarking}

We tested our protocol for three different samples across two measurement techniques. 
First, we investigated a commercially available bulk diamond sample with clear NV ZPLs under a confocal microscope, representing the simple case.
Second, we suspended an ensemble of 20~nm FNDs in DI water and analyzed them using our in-solution setup.
Finally, we analyzed 100~nm FND particles under a confocal microscope. 
For each test cases, we compared our protocol against three other methods for determining the charge state ratio: ZPL fitting, fitting a skewed Gaussian (SG) model, and NNMF (see SI for details). 

\begin{figure*}
    \centering
    \includegraphics[width=0.96\textwidth]{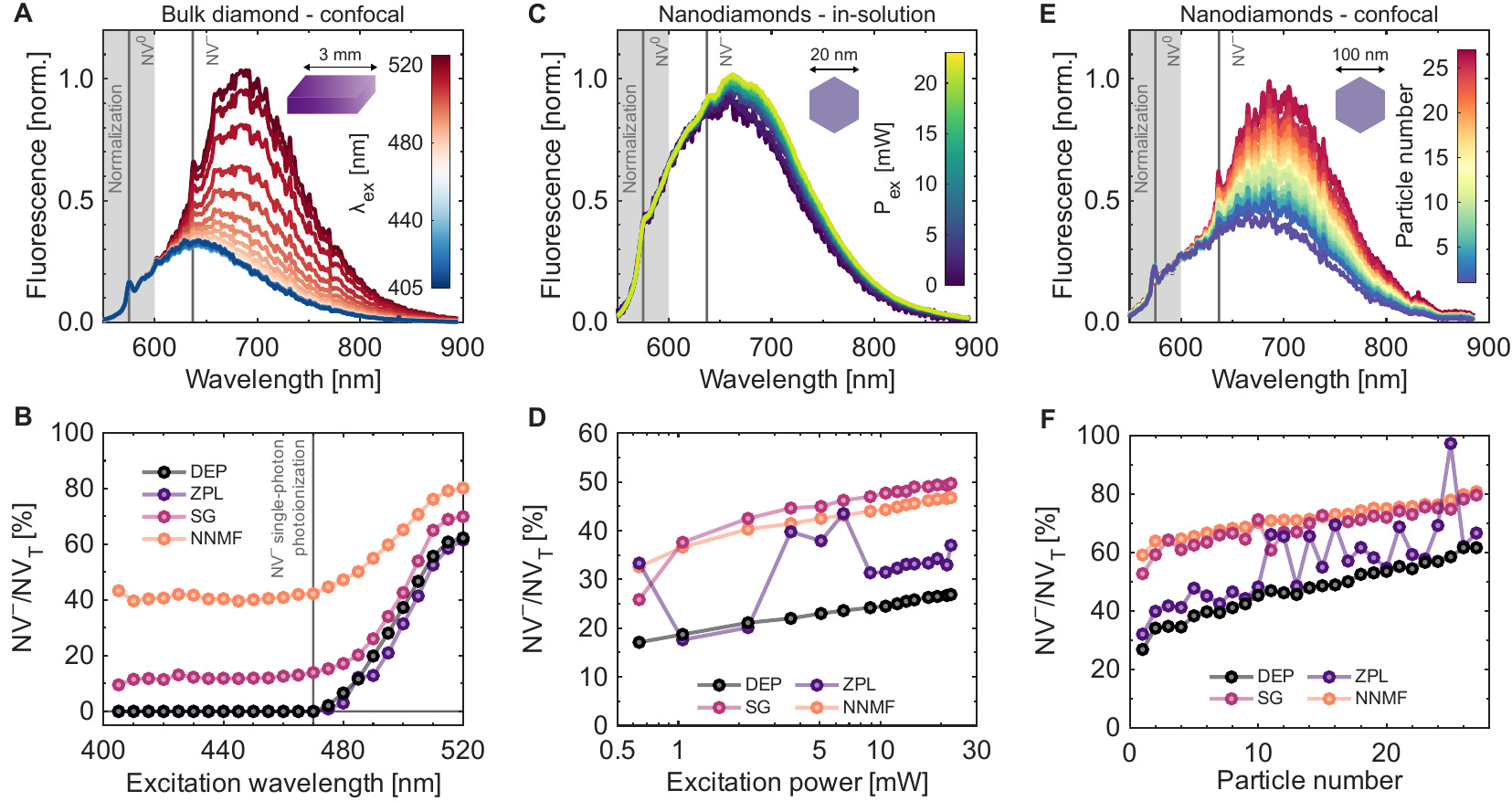}
    \caption{NV charge state ratio protocol comparison.
    \textbf{A:}~Fluorescence spectra of bulk diamond for excitation wavelengths (\lex) ranging \numrange{405}{520}~nm under a confocal microscope.
    \textbf{B:}~\nvm{} contribution of the bulk diamond using our dual-excitation protocol (DEP), ZPL, Skewed Gaussian (SG), and NNMF fitting. 
    The vertical line roughly indicates the wavelength we expect to see single-photon photoionization of \nvm. 
    \textbf{C:}~Fluorescence spectra of 20~nm FNDs suspended in DI water, excited using 520~nm for excitation powers ($P_{\text{ex}}$) between \numrange{0.64}{22.30}~mW.
    \textbf{D:}~\nvm{} contribution of the 20~nm FNDs as a function of excitation power.
    \textbf{E:}~Fluorescence spectra of dispersed 100~nm FNDs under a confocal microscope, ordered by increasing \nvm{} content. 
    \textbf{F:}~\nvm{} contribution for all 100~nm particles using each protocol. 
    All spectra (A, C, E) were normalized between \numrange{550}{600}~nm (shaded regions), with vertical lines marking the positions of the NV ZPLs.
    }
    \label{fig:protocol_testing}
\end{figure*}

We excited the bulk sample using a tunable laser from \numrange{405}{520}~nm (bandwidth of 10~nm) at a constant power of 5~\si{\uW} using a custom-built confocal microscope. 
Normalized fluorescence spectra acquired using different excitation wavelengths are shown in \cref{fig:protocol_testing}A. 
The spectra show mostly \nvn{} PL for excitation wavelengths \numrange{405}{470}~nm. 
For longer wavelengths, the \nvm{} contribution continuously increases up to the highest investigated excitation wavelength of 520 nm. 
The spectra were then analyzed using the different methods mentioned above to determine the \nvm{} ratio as a function of excitation wavelength (\cref{fig:protocol_testing}B, for fitting of individual spectra using each technique, see SI fig. S2-S5). 
All protocols yield a constant NV ratio below 470~nm excitation, which then increases with excitation wavelength up to 520~nm excitation. 
DEP, NNMF, and SG fitting were all capable of fitting each combination of \nvn{} and \nvm{} with a $R^2$ of $>0.98$, with NNMF being the highest on average (SI fig. S6). Both DEP and ZPL fitting yield an \nvm{} ratio of 0~\% below 470~nm excitation, which increases to 60~\% as the excitation wavelength increases to 520~nm.
SG and NNMF on the other hand yield a significantly higher minimum \nvm{} contribution of ~10~\% and 40~\%, respectively, below 470~nm excitation, which increases to 70~\% and 80~\% for 520~nm excitation. 

Assuming that excitation below 450~nm should result in pure \nvn{} fluorescence~\cite{beha2012optimum,manson2018nv,aslam2013photoinduced}, the presence of \nvm{} in these fits is indicative of an underestimate of how far into the red spectral region the \nvn{} emission extends. 
Indeed, the NNMF output for the pure \nvn{} spectrum (SI fig. S7), suggests that \nvn{} does not emit above 750~nm, while our measurements using 400~nm excitation suggest that \nvn{} emission extends well above 800~nm. 
Overall, the data in \cref{fig:protocol_testing}A and B show that both DEP and ZPL fitting yield very similar results in this case, both of which are in agreement with our current understanding of NV charge state photophysics. 
Both NNMF and SG methods, even in this relatively simple case, overestimate the \nvm{} contribution to the NV PL spectra. 

As a second, more challenging test case, we investigated an ensemble of commercially available 20~nm FNDs suspended in water.
FNDs of this size typically show predominantly \nvn{} PL and the ZPLs of both charge states are weak and, in some cases, not visible at all~\cite{eldemrdash2023fluorescent,wilson2019effect}. 
\Cref{fig:protocol_testing}C shows normalized PL spectra of the 20~nm FNDs in suspension as a function of 520~nm laser excitation power in the range from 0.6 to 22~mW. 
The PL spectra slightly red-shift with increasing excitation power, 
which suggests an increase in the \nvm{} contribution with increasing excitation power caused by increased photocycling rates.

We again used the four different analysis approaches to quantify the above observations, with results shown in \cref{fig:protocol_testing}D. 
DEP yields a monotonic increase in the \nvm{} ratio from 17~\% to 27~\% with increasing excitation power. 
While SG and NNMF  yield a comparable relative increase in the \nvm{} contribution with increasing excitation power, both appear to overestimate the absolute \nvm{} ratio by up to 24~\%. 
ZPL fitting yields inconsistent results, though these are typically bounded by the values obtained from the other analysis techniques. 
ZPL fitting is inconsistent mainly because peak fitting becomes extremely sensitive to minor changes in both signal and noise for weak ZPLs such as those observed in 20~nm FNDs. 
Given the relative importance of near-surface NV centers in bulk diamond and FNDs in current research, the ability of DEP to quantify small changes in the \nvm{} ratio even in the absence of strong ZPLs is a major strength of our technique. 

Lastly, we tested our approach for 100~nm FNDs dispersed on a silicon wafer substrate, imaged with a scanning confocal fluorescence microscope connected to a spectrometer. 
We collected PL spectra for 27 FNDs (or small FND aggregates) using 405 and 520~nm excitation (the fluorescence map with marked FNDs can be found in SI fig. S8). 
We averaged all spectra acquired using 405~nm excitation to be used as the \nvn{} reference for DEP. 
The PL spectrum for each particle was normalized and ordered by increasing \nvm{} PL intensity, quantified as the integrated fluorescence above 650~nm divided by the integrated fluorescence below 650~nm, shown in \cref{fig:protocol_testing}E.
Here, we see significant variability in the NV charge state ratio between individual particles, with some consisting almost entirely of \nvn{} while others show predominately \nvm, consistent with prior reports~\cite{reineck2019not}.
\Cref{fig:protocol_testing}F shows the \nvm{} ratio determined by all four analysis approaches for all particles. As observed in the previous two test cases, both SG and NNMF yield a higher \nvm{} contribution, ranging from $\approx$\numrange{55}{80}~\%, compared to both ZPL and DEP (\cref{fig:protocol_testing}F). 
DEP suggests the \nvm{} contribution ranges from \numrange{27}{62}~\% with an average of \num{47(9)}~\%. 
For many particles, ZPL fitting and DEP yield very similar results. 
However, since the ZPL lines are very pronounced for some particles and are barely visible for others, the \nvm{} ratios determined using the ZPL fitting approach are quite susceptible to artefacts and, consequently, exhibit far greater variation. 
We do not, therefore, attribute the observed variation to intrinsic properties of the analysed particles. 

\subsection{Workflow}
\begin{figure}
\centering
\includegraphics[width=0.48\textwidth]{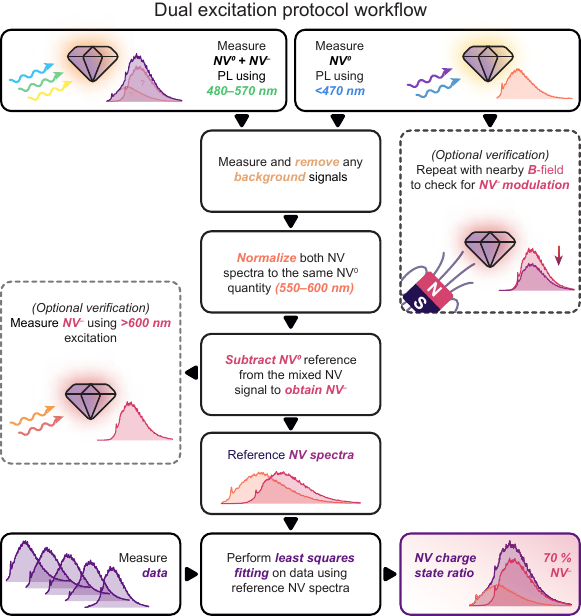}
\caption{Dual excitation protocol workflow.
DEP requires two initial measurements taken with $<470$~nm and \numrange{480}{570}~nm to obtain reference \nvn{} and mixed \nvn/\nvm{} spectra.
Background signals must be removed from the two measurements before normalization between \numrange{550}{600}~nm. 
Once normalized, the difference between the signals can be calculated to obtain the \nvm{} reference spectrum.
Once both reference spectra are determined, they can be used to fit data to determine the relative contribution of each charge state.
}
\label{fig_workflow}
\end{figure}

We propose a simple workflow for the implementation of our protocol, which is summarized in \cref{fig_workflow}. 
Only two primary measurements (\cref{fig_workflow}, top row) are required for the determination of pure \nvm{} and \nvn{} reference spectra; one PL spectrum acquired using green (480-570 nm) excitation that contains contributions from both charge states, and one PL spectrum acquired using blue ($<$470~nm) excitation for a pure \nvn{} spectrum. 
Next, all background signals must be acquired and removed from the two reference spectra. 
Background signals are system and sample dependent, but in many cases this step may only require a simple background subtraction due to detector dark counts. In some cases however, background signals may be more complex, such as Raman signals from the diamond host or the surrounding media (e.g., water) in the case of FND suspensions.

Generally, the acquisition and subtraction of these background spectra is very effective, and we present a complex example of such background removal in SI fig. S9.  
Once all background signals are accounted for, the two spectra (\nvn{}+\nvm{} and \nvn{} only) are normalized between \numrange{550}{600}~nm to the same \nvn{} contribution. Then, the \nvn{} spectrum is subtracted from the spectrum containing both charge states to yield a pure \nvm{} reference spectrum. Finally, the two reference spectra can be fit to data of interest using a least squares fitting routine to determine the relative contribution of each NV charge state to the overall fluorescence signal.

\subsection{Verification and refinement}
Two optional steps can be introduced to verify the NV reference spectra. 
Firstly, it is possible under certain conditions to observe \nvm{} PL using photon energies above 2.63~eV (below 470~nm) which would normally photoionize \nvm{} and create \nvn. 
Therefore, an additional step may be introduced to check for \nvm{} modulation under $<$470~nm excitation in an external magnetic field, which is known to affect the \nvm{} quantum yield~\cite{capelli2017magnetic,tetienne2012magneticfielddependent}, but not \nvn.
Secondly, an excitation wavelength of $>$600~nm can be used to generate a pure \nvm{} spectrum. 
Generally, optical notch filters would be required to remove the excitation light signal from the PL spectrum, blocking part of the NV spectrum. 
Nevertheless, the resulting NV spectrum can be used to optimize the normalization window before the \nvn{} reference is subtracted from the mixed \nvn/\nvm{} signal, or used to remove any remaining \nvm{} in the \nvn{} reference.

\begin{figure}
\centering
\includegraphics[width=0.48\textwidth]{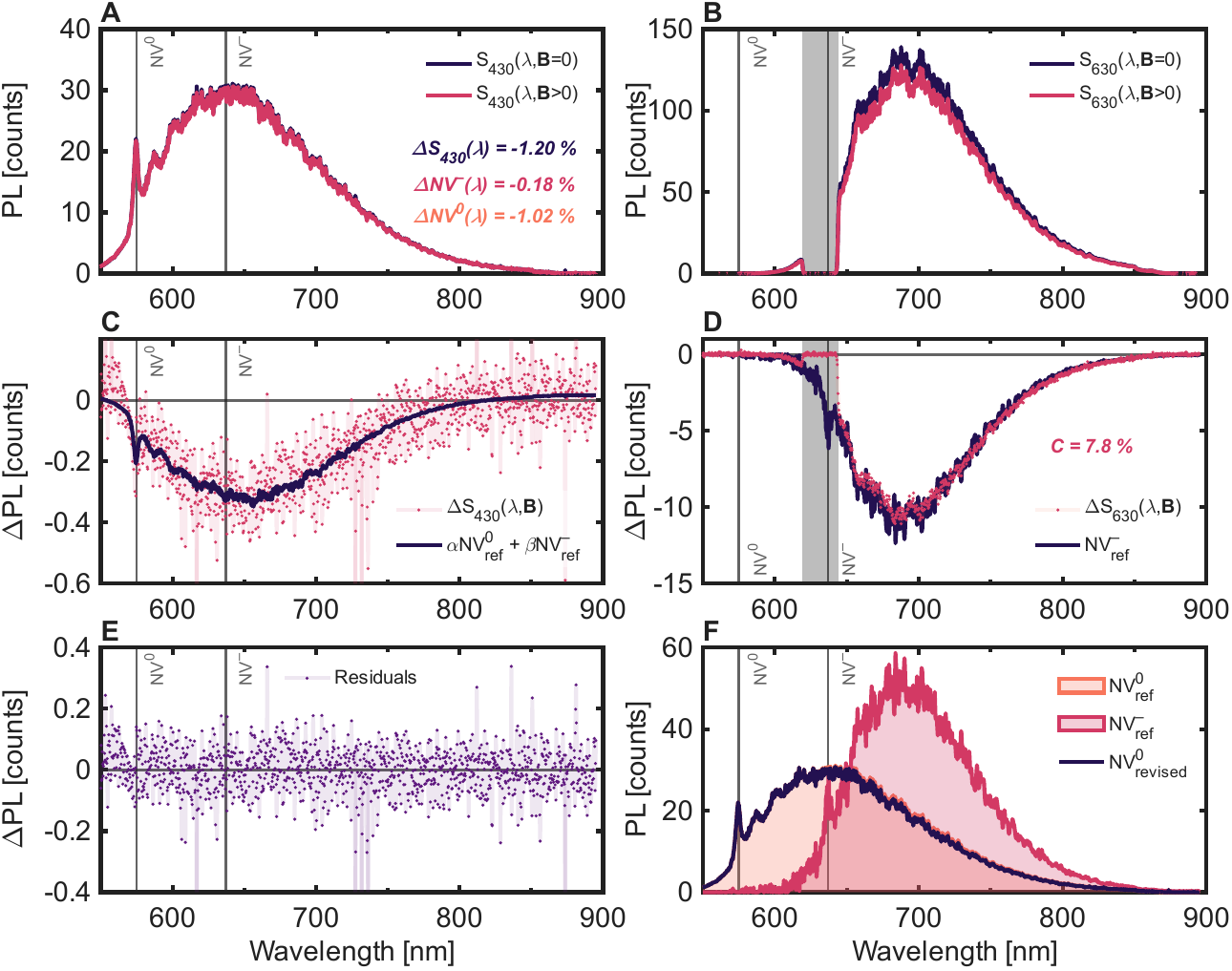}
\caption{NV fluorescence spectra measured using an excitation wavelength of 430~nm (A) and 630~nm (B) with and without the presence of an external $\vec{B}$-field, with the difference shown in (C) and (D), respectively.
The change in fluorescence under 430~nm is fit with the NV reference spectra, with residuals shown in (E).
F: original and revised NV reference spectra.}
\label{fig:nv_verification}
\end{figure}

Here, we measure the NV fluorescence of a bulk diamond sample under both blue (430~nm) and red (630~nm) excitation wavelengths (\cref{fig:nv_verification}A and B, respectively) with and without the presence of an external magnetic field applied via permanent magnet.
Under blue excitation, \cref{fig:nv_verification}C shows a 1.20~\% reduction in overall fluorescence when the magnet is introduced.
Fitting the NV reference spectra using \cref{eq:nv_fitting} allows us to approximate the change in fluorescence attributable to \nvm, which we find to be on order 0.18~\% (residuals in \cref{fig:nv_verification}E).
We show that \nvm{} is reduced by 7.8~\% under the same applied $\vec{B}$-field, which we determine through red (630~nm) excitation and display in \cref{fig:nv_verification}D. 
Under this excitation condition, the change in fluorescence can be attributed solely to spin mixing induced by the magnetic field. 
Assuming that the total \nvm{} is also modulated to the same degree under 430~nm excitation, we can determine how much \nvm{} exists under 430~nm excitation via

\begin{equation}\label{eq:b-fieldcomp}
    \beta_0 = \frac{\Delta\beta}{C}
\end{equation}

where $\beta_0$ is the \nvm{} scaling factor under no magnetic field, $\Delta\beta$ is the change in the \nvm{} scaling factor after the magnet is introduced, and $C$ is the contrast observed under 630~nm (see \cref{fig:nv_verification}D).
By approximating $\beta_0$, we can subtract any remaining \nvm{} from the \nvn{} reference via

\begin{equation}\label{eq:refine_nv0}
    \text{NV}^0_\text{revised} = \text{NV}^0_\text{ref} - \beta_0 \text{NV}^-,
\end{equation}

to obtain a revised \nvn{} reference spectrum which is $\approx2$~\% different to our original approximation as shown in \cref{fig:nv_verification}F.
We note that this correction is still an approximation, as the the fit used to determine $\Delta\beta$ uses an \nvn{} reference which may contain some small fraction of \nvm{} and thus lead to a slight underestimate of $\Delta\beta$. 
Nevertheless, the change is relatively small, which increases our confidence that blue excitation is sufficient for NV charge state determination.

\section{Conclusion}
We investigated several approaches for calculating the \nvn/\nvm{} charge state ratio, particularly for diamond samples which do not exhibit pronounced zero-phonon lines (ZPLs), such as nanodiamonds. 
We tested ZPL fitting, two algorithmic approaches, namely fitting a skewed Gaussian model of the \nvn/\nvm{} spectral profiles and non-negative matrix factorization, and our dual excitation protocol (DEP), which employs blue (\numrange{400}{470}~nm) and green (\numrange{480}{570}~nm) excitation to derive a set of sample- and setup-specific pure \nvn/\nvm{} spectra to fit subsequent experiments. 
We evaluated each approach for a variety of different samples and optical systems, including bulk diamond and dispersed 100~nm nanodiamonds under a scanning confocal microscope, and 20~nm nanodiamonds in aqueous suspension. 
We demonstrate that DEP is capable of deriving quantitative NV charge state ratios that are consistent with our understanding of NV photophysics across common sample types and measurement setups. 

\section*{Acknowledgements}
We acknowledge the support of the Australian Research Council (ARC) through a Discovery Project (DP220102518) and the ARC Centre of Excellence in Quantum Biotechnology (CE230100021). 
GT acknowledges support through an ARC Linkage grant (LP210300230) in collaboration with Diamond Defence Pty Ltd. 
DJM acknowledges support through a University of Melbourne McKenzie Fellowship. 
ND acknowledges support through a Department of Defense NSDIRG grant (NS220100071).
PR acknowledges support through an ARC DECRA Fellowship (DE200100279) and an RMIT University Vice-Chancellor’s Senior Research Fellowship. 
This work was carried out in part at RMIT University’s Microscopy \& Microanalysis Facility, a linked laboratory of Microscopy Australia enabled by the National Collaborative Research Infrastructure Strategy (NCRIS). 

\section*{Conflict of Interest Statement}
\noindent The authors have no conflicts to disclose.


\section*{Data Availability Statement}
\noindent Data available on request from the authors.

\section*{References}
\bibliography{bibliography}

\end{document}


\title{Supplementary information \\
Robust quantification of the diamond nitrogen-vacancy center charge state via photoluminescence spectroscopy}

\author{G. Thalassinos}
\affiliation{School of Science, RMIT University, Melbourne, VIC 3001, Australia}

\author{D. J. McCloskey\textsuperscript}
\affiliation{School of Physics, University of Melbourne, Parkville, VIC 3010, Australia}
\affiliation{Australian Research Council Centre of Excellence in Quantum Biotechnology, School of Physics, University of Melbourne, Parkville, VIC 3010, Australia}

\author{A. Mameli}
\affiliation{School of Science, RMIT University, Melbourne, VIC 3001, Australia}

\author{A. J. Healey}
\affiliation{School of Science, RMIT University, Melbourne, VIC 3001, Australia}

\author{C. Pattinson}
\affiliation{School of Physics, University of Melbourne, Parkville, VIC 3010, Australia}
\affiliation{Australian Research Council Centre of Excellence in Quantum Biotechnology, School of Physics, University of Melbourne, Parkville, VIC 3010, Australia}

\author{D. Simpson}
\affiliation{School of Physics, University of Melbourne, Parkville, VIC 3010, Australia}
\affiliation{Australian Research Council Centre of Excellence in Quantum Biotechnology, School of Physics, University of Melbourne, Parkville, VIC 3010, Australia}

\author{B. C. Gibson}
\affiliation{School of Science, RMIT University, Melbourne, VIC 3001, Australia}

\author{A. Stacey}
\affiliation{School of Science, RMIT University, Melbourne, VIC 3001, Australia}

\author{N. Dontschuk}
\affiliation{School of Physics, University of Melbourne, Parkville, VIC 3010, Australia}

\author{P. Reineck}
\affiliation{School of Science, RMIT University, Melbourne, VIC 3001, Australia}

\maketitle

\renewcommand{\thefigure}{S\arabic{figure}}
\renewcommand{\thetable}{S\arabic{table}}

\section{NV emission ranges}
\noindent \Cref{tab:nv_max_emission} provides a summary of the approximate maximum emission wavelength of both \nvn{} and \nvm{} observed across the literature. 
Data is organised as a function of diamond type and size, including excitation wavelength(s) (\lex) used in each study to investigate \nvn{} and/or \nvm. 

\begin{table*}
    \caption{Maximum emission wavelength as measured/predicted by different studies. 
    Boldface denotes \nvn/\nvm{} spectral profiles derived using NNMF.}
    \label{tab:nv_max_emission}
    \renewcommand{\arraystretch}{1.2} 
\begin{ruledtabular}
\begin{tabular}{lllddr}
		\multirow{2}{*}{Author(s)}	& \multirow{2}{*}{Diamond type/size} & \multirow{2}{*}{\lex{} [nm]}	& \multicolumn{2}{c}{Max. emission wavelength [nm]} 	&	\multirow{2}{*}{Ref.}	\\ \cmidrule{4-5}
		~ 		& ~ & ~ 	& \mbox{\nvn}  & \mbox{\nvm}	& ~ \\ \hline
        Gruber et al.       & Type Ib  & 514  & \mbox{---} & 900 & \onlinecite{gruber1997scanning} \\
        Jeske et al.        & Type Ib  & 53  & 830 & 870 & \onlinecite{jeske2017stimulated} \\
        Kurtsiefer et al.   & Type Ib  & 532 & \mbox{---} & 760 & \onlinecite{kurtsiefer2000stable} \\
        Manson et al.       & Type Ib  & 445 (\nvn), 532 (\nvm) & 850   & 850 & \onlinecite{manson2018nv} \\
        Manson and Harrison & Type Ib  & 532  & \mbox{$>$}740 & \mbox{$>$}740 & \onlinecite{manson2005photoionization} \\
        \textbf{Savinov et al.}      & \textbf{Type Ib}  & \textbf{532}  & \textbf{740} & \textbf{\mbox{$>$}800} & \onlinecite{savinov2022diamond} \\
        Alsid et al.        & Type IIa & 532  & \mbox{$>$}800    & \mbox{$>$}800 & \onlinecite{alsid2019photoluminescence} \\   
        Beha et al.         & Type IIa & 521, 550  & \mbox{---}       & \mbox{$<$}800 & \onlinecite{beha2012optimum} \\
        Groot-Berning et al. & Type IIa     & 532  & 850   & 850 & \onlinecite{groot-berning2014passive} \\
        Aslam et al.        & Type IIa & 440 (\nvn), 514 (\nvm)  & 850  & 850   & \onlinecite{aslam2013photoinduced} \\
        Aude Craik et al.   & Type IIa & 532                     & 850  & 850   & \onlinecite{audecraik2020microwaveassisted} \\
        Dhomkar et al.      & Type IIa & 532  & \mbox{---}   & 800 & \onlinecite{dhomkar2018charge} \\
        \textbf{McCloskey et al.}    & \textbf{Type IIa} & \textbf{532} & \textbf{775} & \textbf{850} & \onlinecite{mccloskey2020enhanced} \\
        Bhaumik et al.      & \mbox{---}      & 532  & \mbox{$>$}750 & \mbox{$>$}750 & \onlinecite{bhaumik2019tunable} \\
        \textbf{Reineck et al.}      & \textbf{HPHT ND, 100--200~nm} & \textbf{520} & \textbf{720} & \textbf{790} & \onlinecite{reineck2019not} \\
        \textbf{Wilson et al.}       & \textbf{HPHT ND, 10--140~nm} & \textbf{510} & \textbf{725} & \textbf{850} & \onlinecite{wilson2019effect} \\
        Mohtashami and Koenderink & HPHT ND, 26--108~nm & 532 & \mbox{---} & 850 & \onlinecite{mohtashami2013suitability}  \\
        Rondin et al.       & HPHT ND, 10--100~nm & 532 & \mbox{$<$}800 & \mbox{$<$}800 & \onlinecite{rondin2010surfaceinduced} \\
        Treussart et al.    & HPHT ND, 50~nm & 514.5 & \mbox{$<$}800 & \mbox{$<$}800 & \onlinecite{treussart2006photoluminescence}  \\
        Faklaris et al.     & HPHT ND, \mbox{$<$}50 nm & 532 & \mbox{---} & \mbox{$<$}800 & \onlinecite{faklaris2010photoluminescent} \\
        Faklaris et al.     & HPHT ND, \mbox{$<$}50~nm & 488 & 720  & \mbox{---} & \onlinecite{faklaris2009photoluminescent} \\
        Berthel et al.      & HPHT ND, 25~nm & 515 & \mbox{---} & \mbox{$>$}800 & \onlinecite{berthel2015photophysics} \\
        Karaveli et al.     & HPHT ND, \mbox{$<$}20~nm  & 532 & 800   & 800 & \onlinecite{karaveli2016modulation} \\
        Bradac et al.       & DND, 5~nm & 532 & \mbox{---} & 840 & \onlinecite{bradac2010observation} \\
\end{tabular}
\end{ruledtabular}
\end{table*}

Generally, the maximum emission range of both NV charge states vary significantly throughout the literature, with no clear trend emerging as a function of diamond type or size. 
Works investigating NV fluorescence using blue (\numrange{400}{450}~nm) excitation show negligible difference in maximum emission wavelengths between \nvn{} and \nvm~\cite{manson2018nv,aslam2013photoinduced}.
A similar observation can be made for works which derive ``pure'' NV spectra experimentally~\cite{audecraik2020microwaveassisted,alsid2019photoluminescence}, which is contrasting against works using non-negative matrix factorization (NNMF), which suggests that \nvn{} emits far less into the red spectrum compared to \nvm~\cite{savinov2022diamond,mccloskey2020enhanced,reineck2019not,wilson2019effect}.

\newpage
\section{Samples}
\noindent \Cref{tab:samples} provides a summary of samples used throughout this work and which figures they appear in.

\begin{table}[h]
    \centering
    \caption{Summary of investigated samples.}
    \label{tab:samples}
    \renewcommand{\arraystretch}{1.5} 
    \begin{ruledtabular}
    \begin{tabular}{lccl}
        \textbf{Sample} & \textbf{System} & \textbf{Shown in figure(s)} & \textbf{Manufacturer} \\
    \hline
        Bulk DNV-B14 & Confocal & 1A, 3A-B, 5A-F, S2-S7 & Element Six \\
        140 nm HPHT FND & Solution & 2A-C & Adàmas Nano. \\
        100 nm HPHT FND & Confocal & 3E-F, S9 & Adàmas Nano. \\
        20 nm HPHT FND & Solution & 3C-D, S8 & Adàmas Nano. \\

\end{tabular}
\end{ruledtabular}
\end{table}

\section{Optical setups}
\begin{figure*}[tbh]
    \centering
    \includegraphics[width=1\linewidth]{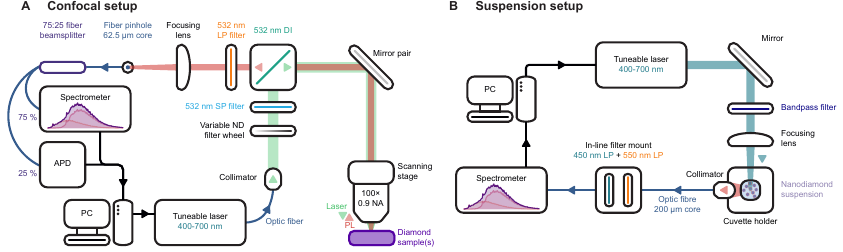}
    \caption{Schematics of custom built confocal (A) and suspension (B) setups.
    \textbf{A:}~A fibre coupled, pulsed tuneable laser is sent through a 532~nm SP filter and variable ND filter to control excitation power and guided to the sample through a 532~nm DI mirror and a pair of mirrors. An XYZ scanning stage with a 100$\times$ objective (NA = 0.9) probes the diamond sample(s). Fluorescence is collected through the objective and is focused onto a multi-mode fibre with a in-built beamsplitter, which sends 75~\% of the signal to the spectrometer and the remaining 25~\% to an APD.
    \textbf{B:}~A tuneable laser is reflected off a mirror, through a bandpass filter and focused onto a cuvette holding the nanodiamond suspension. Fluorescence is collected through a multi-mode fibre, sent through an in-line filter mount, then coupled to a spectrometer. 
    }
    \label{fig:optical_setups}
\end{figure*}

Schematics of our custom-built scanning confocal microscope and in-solution spectroscopy setup can be seen in \cref{fig:optical_setups}.

\newpage
\section{NV charge state determination protocols}
\subsection{Dual excitation protocol}
\begin{figure*}[h]
    \centering
    \includegraphics[width=1\linewidth]{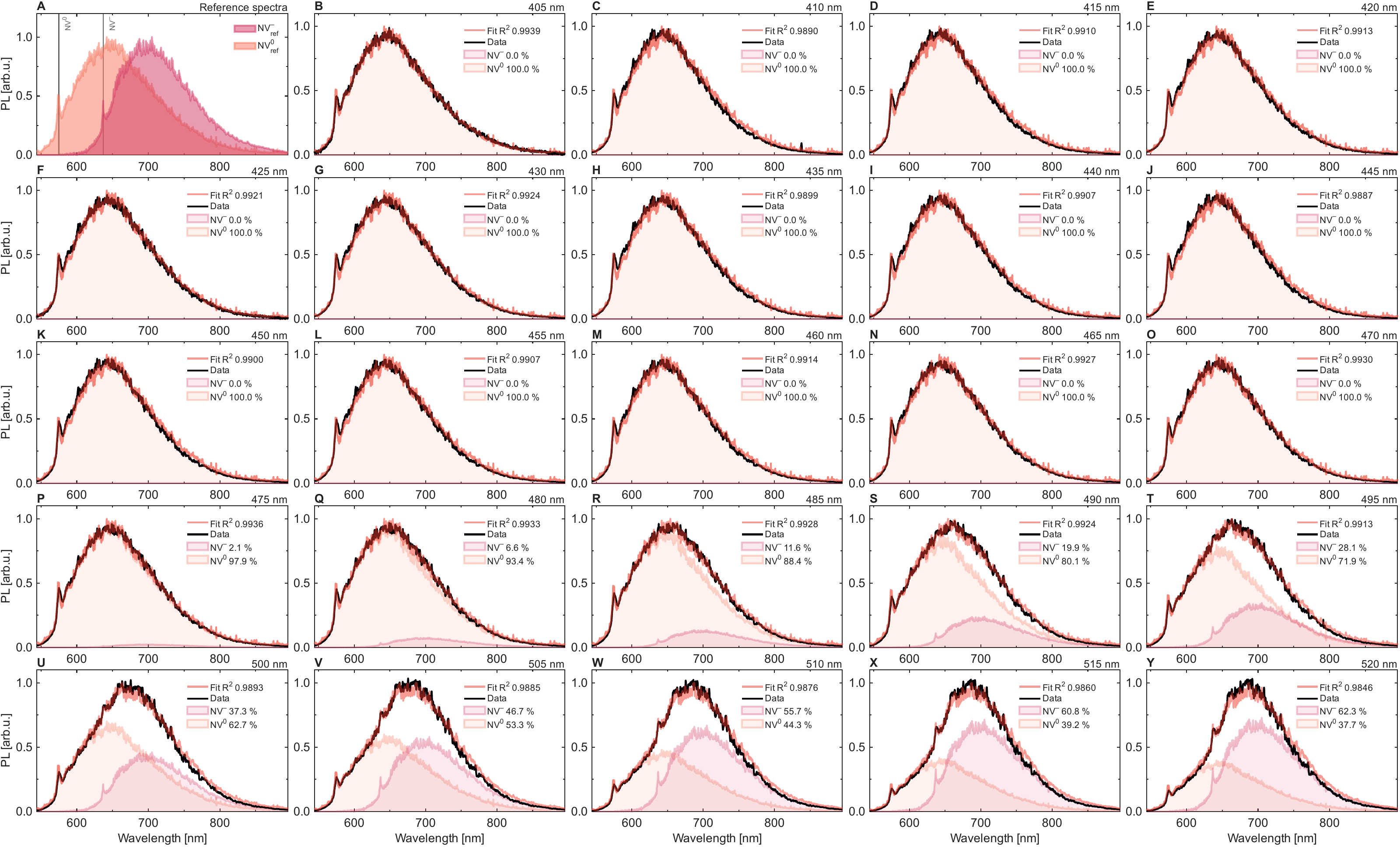}
    \caption{NV charge state analysis via dual excitation protocol (DEP) on DNV-B14.
    \textbf{A:}~Predicted pure \nvn{} and \nvm{} spectra based on DEP.
    \textbf{B-Y:}~Fitting of DEP NV spectra to fluorescence spectra taken for excitation wavelengths between \numrange{405}{520}~nm in increments of 5~nm and a laser bandwidth of 10~nm.
    }
    \label{fig:si:B14_ChargeStateAnalysis_DEP}
\end{figure*}

\noindent Here, we show individual fits using each of the NV charge state ratio determination techniques used to produce the data in figure 3B of the main text. 
\Cref{fig:si:B14_ChargeStateAnalysis_DEP} shows our dual excitation protocol (DEP) on bulk diamond across excitation wavelengths between \numrange{405}{520}~nm, with reference \nvn/\nvm{} spectra shown in \Cref{fig:si:B14_ChargeStateAnalysis_DEP}A.

\newpage
\subsection{Skewed Gaussian fitting}
\noindent For skewed Gaussian (SG) fitting, we used a custom fit equation to approximate the NV fluorescence spectra.
We used a combination of two regular Gaussian functions to model both \nvn{} and \nvm{} zero-phonon lines (ZPLs) and two skewed Gaussians for their respective phonon side-bands (PSBs) via

\begin{equation}\label{eq:SG_model}
\begin{split}
    S_\text{NV}(\lambda) =~  
    & A_\text{ZPL1} \cdot \exp\left( - \frac{\left( \lambda-\lambda_\text{ZPL1} \right)^2}{2 \nu_\text{ZPL1}^2} \right) +
      A_\text{PSB1} \cdot \exp\left(-\left( \frac{\lambda-\lambda_\text{PSB1}}{\nu_\text{PSB1}} \right)^2\right) \cdot \left( 1 + \text{erf} \left(\alpha_1 \cdot \frac{\lambda-\lambda_\text{PSB1}}{\nu_\text{PSB1}} \right)\right) + \\
    & A_\text{ZPL2} \cdot \exp\left( - \frac{\left( \lambda-\lambda_\text{ZPL2} \right)^2}{2 \nu_\text{ZPL2}^2} \right) + 
     A_\text{PSB2} \cdot \exp\left(-\left( \frac{\lambda-\lambda_\text{PSB2}}{\nu_\text{PSB2}} \right)^2\right) \cdot \left( 1 + \text{erf} \left(\alpha_2 \cdot \frac{\lambda-\lambda_\text{PSB2}}{\nu_\text{PSB2}}\right)\right),
\end{split}
\end{equation}

where erf is the error function, given by

\begin{equation}
    \text{erf}(x) = \frac{2}{\sqrt{\pi}}\int_0^x e^{-t^2} dt.
\end{equation}

\begin{figure*}[bh]
    \centering
    \includegraphics[width=1\linewidth]{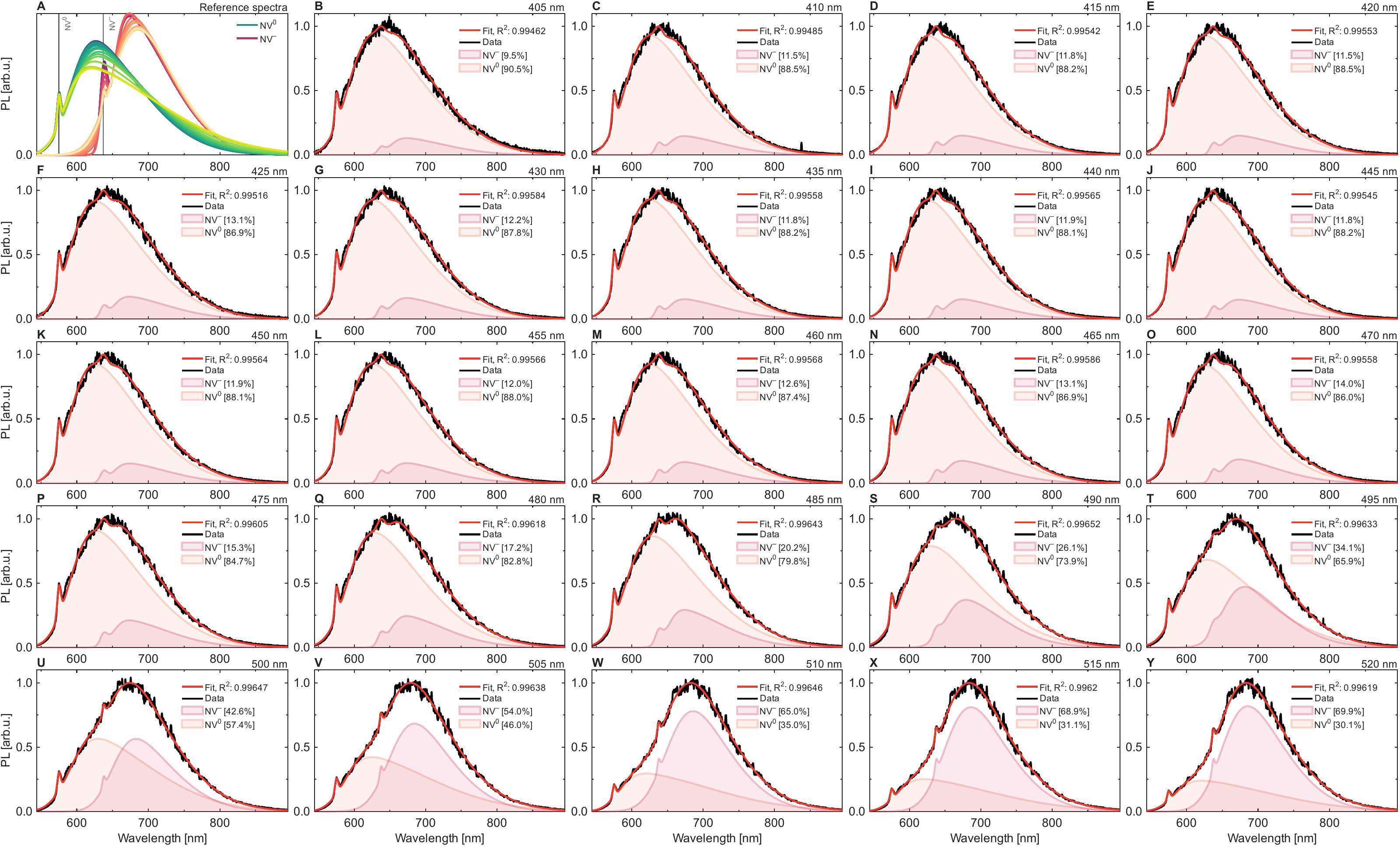}
    \caption{NV charge state analysis via skewed Guassian (SG) fitting on DNV-B14.
    \textbf{A:}~Reference \nvn/\nvm{} spectra used to fit experimental data. Fit parameters differed slightly for each measurement, leading to different NV reference spectra. 
    \textbf{B-Y:}~Fitting of DEP NV spectra to fluorescence spectra taken for excitation wavelengths between \numrange{405}{520}~nm in increments of 5~nm and a laser bandwidth of 10~nm.
    }
    \label{fig:si:B14_ChargeStateAnalysis_SG}
\end{figure*}

We restricted all free parameters to a small window of acceptable values, enough to fit all combinations of \nvn{} and \nvm. 
However, we could not find a set of parameters that would allow one pair of \nvn/\nvm{} reference spectra to fit the full series of data. 
Thus, the reference \nvn/\nvm{} spectra were slightly different for each excitation wavelength. 
These differences can be seen in \cref{fig:si:B14_ChargeStateAnalysis_SG}A, where all \nvn{} and \nvm{} reference spectra are presented together. 
Despite these differences, we were able to fit all experimental data (\cref{fig:si:B14_ChargeStateAnalysis_SG}B--Y) using this technique and measure a systematic change in the NV charge state ratio. 

\newpage
\subsection{Non-negative matrix factorization}
\begin{figure*}[hbt]
    \centering
    \includegraphics[width=1\linewidth]{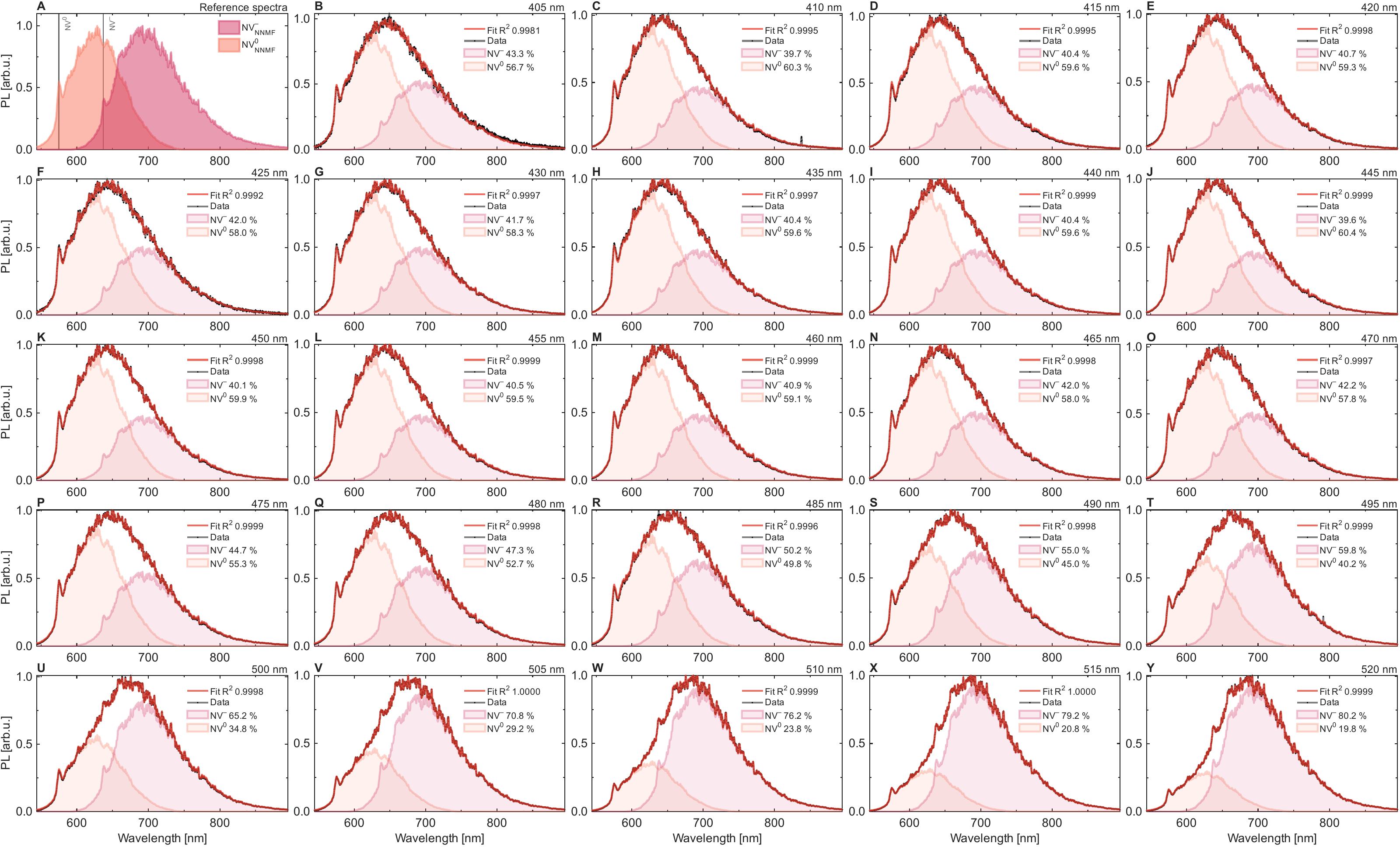}
    \caption{NV charge state analysis via NNMF on DNV-B14.
    \textbf{A:}~Predicted pure \nvn{} and \nvm{} spectra based on NNMF.
    \textbf{B-Y:}~Fitting of NNMF NV spectra to fluorescence spectra taken for excitation wavelengths between \numrange{405}{520}~nm in increments of 5~nm and a laser bandwidth of 10~nm.}
    \label{fig:si:B14_ChargeStateAnalysis_NNMF}
\end{figure*}

\noindent For non-negative matrix factorization (NNMF), we used all background-corrected data for excitation wavelengths \numrange{405}{520}~nm as inputs for the `nnmf' function in MATLAB, assuming 2 components. 
The outputs are shown in \cref{fig:si:B14_ChargeStateAnalysis_NNMF}A, with fits for each measurement across \cref{fig:si:B14_ChargeStateAnalysis_NNMF}B--Y. 

\newpage
\subsection{Zero-phonon line fitting}
\noindent Finally, for our ZPL method, we started by fitting two second order polynomials around both NV ZPLs at 575 and 637~nm, ignoring counts $\pm$5~nm around each ZPL. 
This baseline, which represented fluorescence originating from the PSBs, rather than the ZPLs, was set to 0. 
We then fit a Gaussian (or Lorentzian, sample-dependent) function onto each ZPL and integrated the area under the curve to calculate the \nvn/\nvm{} ZPL intensities ($I_{\text{NV}}$).
The \nvm{} contribution is then given as

\begin{equation}
    \text{NV}^- [\%] = \frac{I_{\text{NV}^-}}{I_{\text{NV}^0} + I_{\text{NV}^-}}.
\end{equation}

\begin{figure*}[h]
    \centering
    \includegraphics[width=1\linewidth]{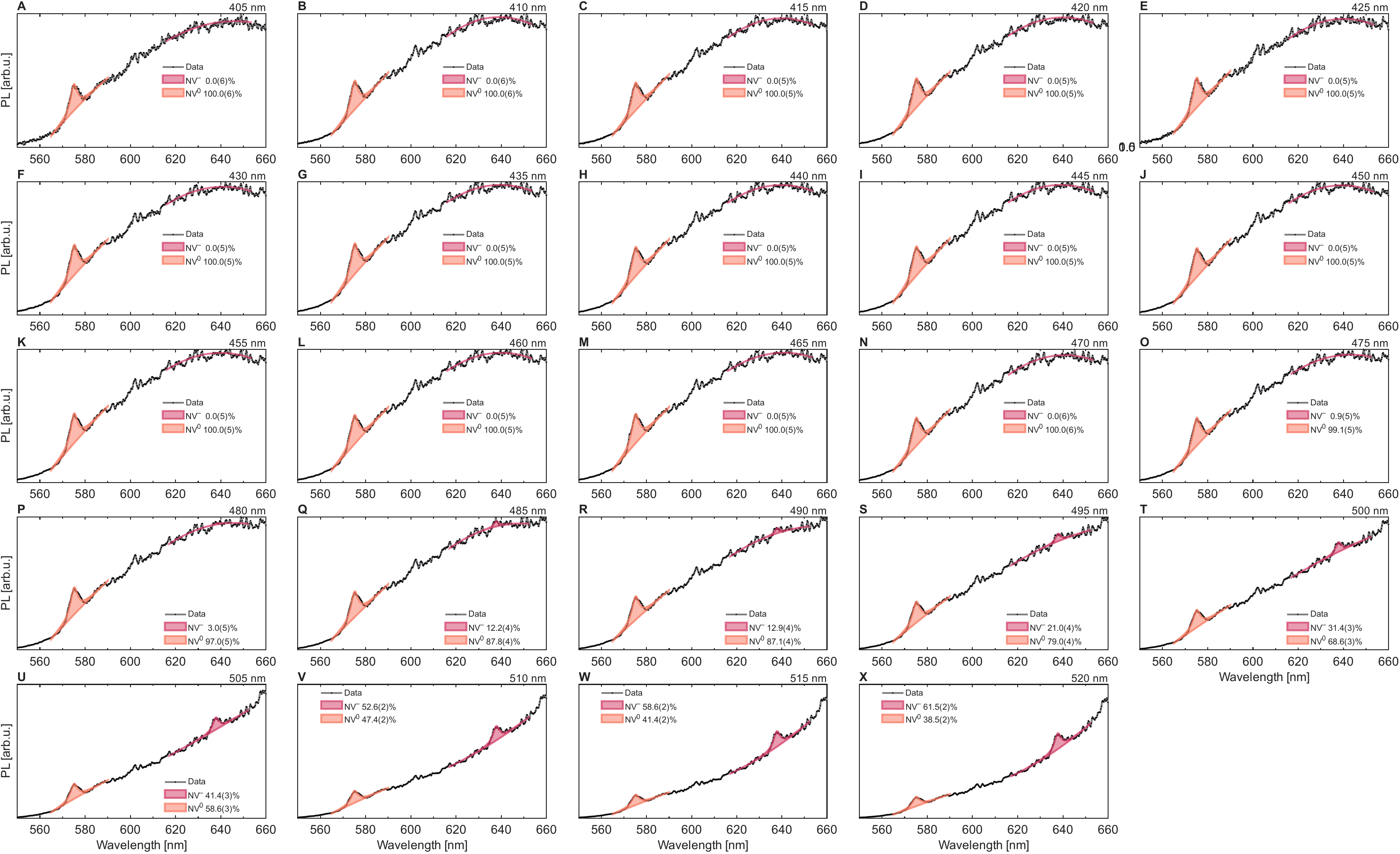}
    \caption{NV charge state analysis via ZPL fitting on DNV-B14.
    \textbf{A-X:}~Fitting of \nvn/\nvm{} ZPL for spectra taken under excitation wavelengths between \numrange{405}{520}~nm in increments of 5~nm and a laser bandwidth of 10~nm.}
    \label{fig:si:B14_ChargeStateAnalysis_ZPL}
\end{figure*}

\Cref{fig:si:B14_ChargeStateAnalysis_ZPL} shows ZPL fitting for each excitation wavelength for the bulk sample. 
The relative contribution of each charge state is shown as part of the legend in each panel (A--X) with values in parentheses denoting uncertainty originating from uncertainties in each fit. 

\newpage
\subsection{Technique validity}
\begin{figure}[bht]
    \centering
    \includegraphics[width=0.5\linewidth]{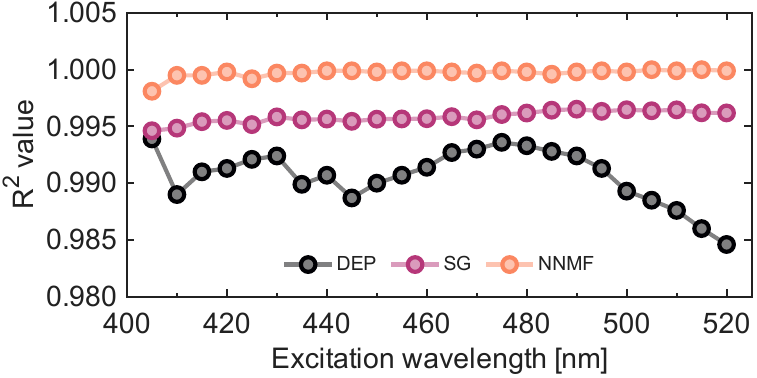}
    \caption{$R^2$ values of all fits corresponding to figures S1--3.}
    \label{fig:si:r2_vals}
\end{figure}

\noindent For DEP, SG, and NNMF, we calculated the $R^2$ values of each individual measurement shown in \cref{fig:si:B14_ChargeStateAnalysis_DEP,fig:si:B14_ChargeStateAnalysis_SG,fig:si:B14_ChargeStateAnalysis_NNMF} to quantify how well each method models the NV fluorescence spectra for different NV charge state ratios, shown in \cref{fig:si:r2_vals}. 
All three techniques show high $R^2$ values across the NV charge state ratio range we observed, with minimum $R^2$ value of 0.9846, and NNMF demonstrating the best fitting across the full range of excitation wavelengths.
However, while both NNMF and SG models fit the experimental data better than our DEP, neither model can be physically justified. 

We show in \cref{fig:si:B14_ChargeStateAnalysis_SG}A that the only way to achieve a good fit using our SG model is to allow fitting parameters to change per measurement. 
This leads to significantly different reference \nvn/\nvm{} spectra depending on the NV charge state ratio. 
However, we know that the fluorescence spectral profiles of both \nvn{} and \nvm{} are fixed quantities for a given sample. 
Therefore, the NV charge state ratio calculated using this technique is unlikely to represent the ``true'' NV charge state ratio. 

In the case of NNMF, which shows the best overall fitting with an average $R^2$ value of 0.9997, we see that the prediction for pure \nvn{} shows emission only up to $\approx$750~nm (\cref{fig:si:B14_ChargeStateAnalysis_NNMF}A), significantly shorter than experimental evidence across the literature (see \cref{tab:nv_max_emission}) and predictions made by other techniques~\cite{alsid2019photoluminescence,audecraik2020microwaveassisted}.

\begin{figure*}
    \centering
    \includegraphics[width=0.9\linewidth]{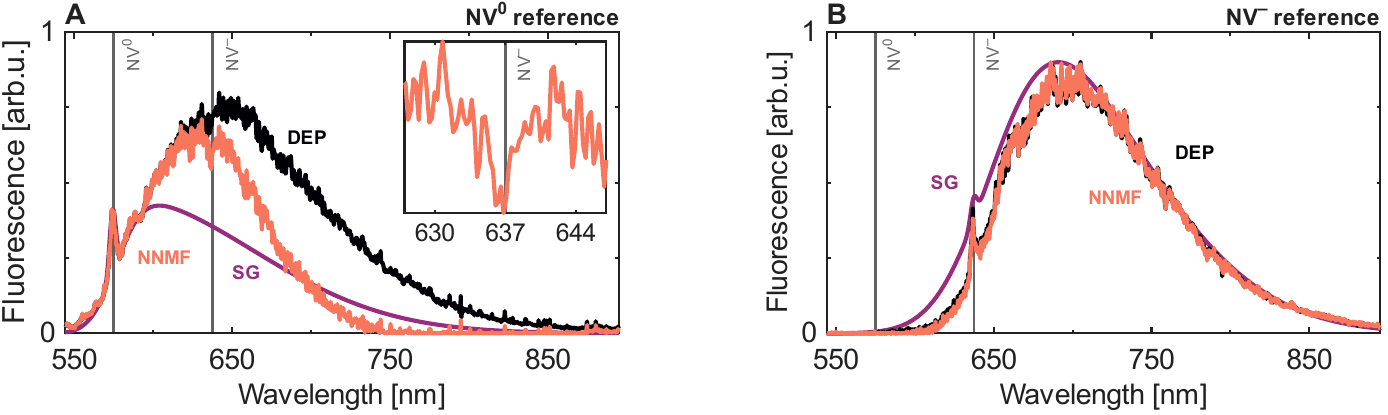}
    \caption{\nvn/\nvm{} spectral profiles of bulk diamond as derived via DEP, NNMF, and SG fitting used to fit spectra in \cref{fig:si:B14_ChargeStateAnalysis_DEP,fig:si:B14_ChargeStateAnalysis_SG,fig:si:B14_ChargeStateAnalysis_NNMF}, for figure 3 of the main text.
    \textbf{A:}~\nvn{} spectral profiles normalized to 575~nm showing significant variation at wavelengths $>$600~nm.
    The inset shows how the NNMF estimate of \nvn{} has a dip at the \nvm{} ZPL, indicating that it may be underestimating \nvn.
    \textbf{B:}~\nvm{} spectral profile which shows DEP and NNMF deriving identical results, while SG fitting results in a close-approximation to \nvm.
    Dashed lines represent the locations of the NV ZPLs.}
    \label{fig:protocol_ref_comp}
\end{figure*}

We show a direct comparison between the \nvn/\nvm{} reference spectra derived from each of the three techniques in \cref{fig:protocol_ref_comp}. 
While each approach predicts similar a \nvm{} spectrum, with NNMF and DEP being indistinguishable, the \nvn{} spectral profiles between the three methods vary significantly, highlighting the fundamental challenge: that \nvn{} is difficult to verify. 

\newpage

\section{Confocal fluorescence map of fluorescent nanodiamonds}
\begin{figure}[tbh]
    \centering
    \includegraphics[width=0.4\linewidth]{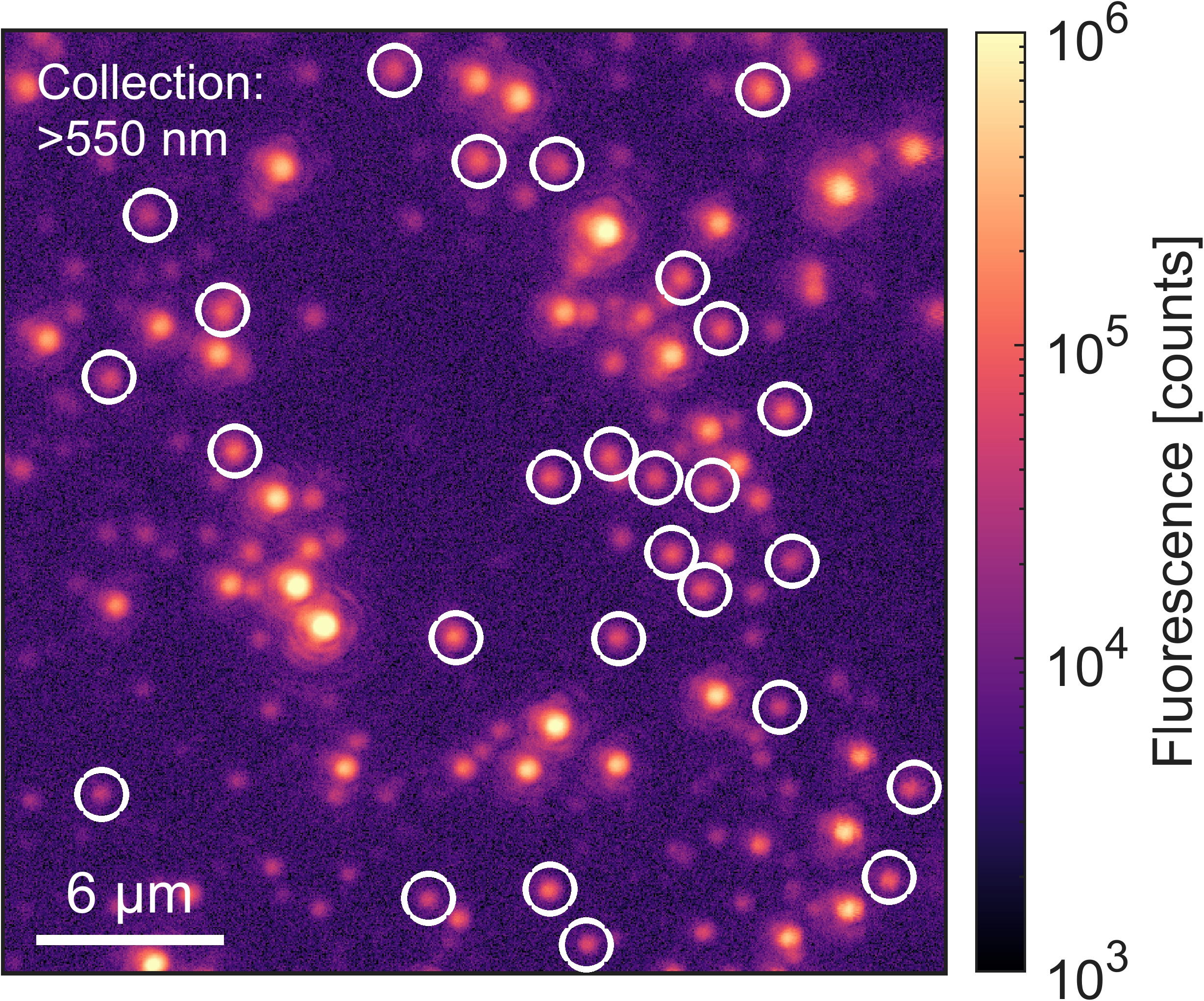}
    \caption{Confocal fluorescence map of 100~nm FND particles spin coated onto a Si wafer substrate excited with 520~nm and and collected through a 550~nm long pass filter.
    White circles mark particles which were analyzed.}
    \label{fig:si:pl_map}
\end{figure}

\noindent \Cref{fig:si:pl_map} shows a confocal fluorescence map of the 100~nm FND particles investigated for figure 3E-F of the main text. 
The fluorescence map was taken under 520~nm excitation with a 550~nm long pass collection filter using our scanning confocal microscope (\cref{fig:optical_setups}A).
The white circles identify particles which were analyzed with our spectrometer.

\section{Background corrections for in-solution measurements}
\noindent Nanoparticles in suspension represent one of the more challenging environments for determining the NV charge state ratio due to the overlap of multiple background signals. 
Under optimal NV excitation conditions of \numrange{510}{540}~nm, the peak of the water Raman spectrum overlaps significantly with NV fluorescence at \numrange{617}{661}~nm.
Furthermore, nanodiamond particles of different sizes and concentrations have different scattering properties, leading to differences in the total intensity of the measured water Raman signal, which is further convoluted by variations in particle brightness. 
Therefore, it is non-trivial to determine how much the water Raman signal contributes to the overall measured signal compared to a pure water reference measurement. 
Under these conditions, there is no available experimental NV reference spectrum which can be used to accurately remove the background signals from the measured spectrum. 

\begin{figure}[tbh]
    \centering
    \includegraphics[width=0.9\linewidth]{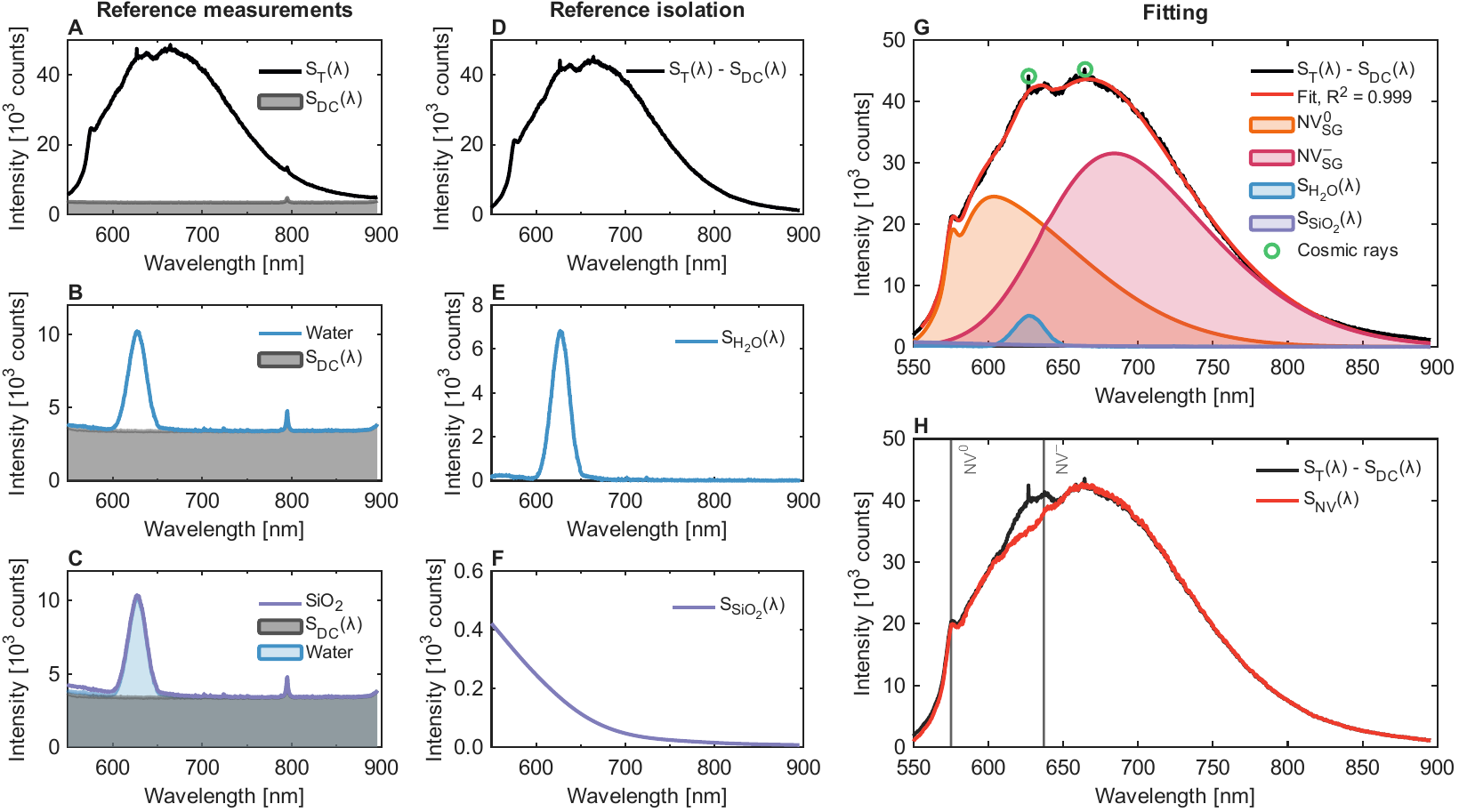}
    \caption{Background correction for in-solution measurements.
    \textbf{A:}~Raw data of nanodiamond fluorescence measured in DI water suspension ($S_T(\lambda)$) and spectrometer dark counts ($S_\text{DC}(\lambda)$). 
    \textbf{B:}~Raw nanodiamond data with dark counts removed.
    \textbf{C:}~Water Raman reference measurement with dark counts.
    \textbf{D:}~Pure water reference spectrum under 520~nm excitation. 
    \textbf{E:}~Measurement of 40~nm SiO$_2$ nanoparticles suspended in DI water.
    \textbf{F:}~Scattering reference with dark counts and water Raman signals removed.
    \textbf{G:}~Fitting of nanodiamond signal using NV SG model (\nvn$_\text{SG}$/\nvm$_\text{SG}$), water, and scattering references, with cosmic rays identified by marked circles.
    \textbf{H:} Comparison before and after the removal of water Raman, scattering, and cosmic ray signals from the nanodiamond sample, showing isolation of NV fluorescence. 
    }
    \label{fig:bkg_correction}
\end{figure}

To precisely remove all background signals and isolate NV fluorescence, we use a combination of multiple background references and skewed Gaussian (SG) fitting. 
In \cref{fig:si:B14_ChargeStateAnalysis_SG}, we show that our SG model (\cref{eq:SG_model}) can be used to fit \nvn/\nvm{} spectra. 
Therefore, we can use this NV fluorescence spectrum model ($S_\text{NV}(\lambda)$) with additional parameters representing the scaling of the background references to fit the raw data ($S_T$) via

\begin{equation}\label{eq:bkg_removal}
    S_T(\lambda) = S_\text{NV}(\lambda) + \kappa_1 S_1(\lambda) + \kappa_2 S_2(\lambda) ... + \kappa_n S_n,
\end{equation}

where $S_n$ and $\kappa_n$ represent the $n^\text{th}$ reference spectrum and it's non-negative scaling, respectively. 
For nanodiamonds in DI water, we recorded the spectrometer dark counts ($S_\text{DC}(\lambda)$), a pure water reference ($S_{\text{H}_2\text{O}}(\lambda)$), and non-fluorescent SiO$_2$ nanoparticles (nominal size 40~nm) suspended in water ($S_{\text{SiO}_2}(\lambda)$) as our background signals. 
A thorough demonstration of our background removal steps is illustrated in \cref{fig:bkg_correction}.

Here, we show each of the measurements (nanodiamonds, dark counts, water, and SiO$_2$ particles) as they were recorded (\cref{fig:bkg_correction}A-C) and with overlapping signals removed (\cref{fig:bkg_correction}D-F). 
For both nanodiamonds and the water Raman reference, we simply subtracted the dark counts. 
However, with the SiO$_2$ particles, we had to also remove the water Raman signal to isolate the scattering reference. 
We then fit the nanodiamond data using \cref{eq:bkg_removal}, with each individual component of the model shown separately as shaded areas, as illustrated in \cref{fig:bkg_correction}G.
We demonstrate that this model fits the nanodiamond data with a high degree of confidence ($R^2 = 0.999$), allowing us to further identify outliers in the signal arising from cosmic rays interacting with our spectrometer. 
We can then remove all background signals, isolating NV fluorescence (\cref{fig:bkg_correction}H). 

\section*{References}
\bibliography{bibliography}